\documentclass[12pt]{article}
\usepackage{amsfonts}
\usepackage{amssymb}
\usepackage{amsmath}
\newcommand{\G}{\Gamma}

\renewcommand{\d}{\delta}
\renewcommand{\b}{\beta}
\newcommand{\g}{\gamma}

\newcommand{\D}{\Delta}

\newcommand{\e}{\epsilon}

\newcommand{\ar}{\longrightarrow}

\newcommand{\la}{\lambda}
\renewcommand{\a}{\alpha}
\begin{document}
\title{About mathematical apparatus of many body quantum dynamics}
\author{Yu.I.Ozhigov\thanks{The work is supported by Fond of NIX Computer Company (grant \# F793/8-05), INTAS grant 04-77-7289 and Russian Fond for Basic Research grant 06-01-00494-a. e-mail: 
ozhigov@cs.msu.su} \\[7mm]
Moscow State University,\\
Institute of Physics and Technology of RAS
} 
\maketitle
\begin{abstract}
We discuss the possibility to modify many-body Hilbert quantum formalism that is necessary for the representation of quantum systems dynamics. The notion of effective classical algorithm and visualization of quantum dynamics play the key role. 
\end{abstract}
\newpage
\section{Introduction}

This work touches the unusual theme - the possibility to modify the mathematical apparatus of quantum theory. Its aim - to show why it is desirable and what future waits quantum theory if this modification comes true. Such a work must bear the general character with necessity, and in this case also its main idea is not conventional. Nevertheless, the discussed theme has the immediate relation to many experiments on new technologies based on quantum effects. Moreover, the hypothesis formulated in this work can be rejected in the real experiments (which are in process), and it thus has completely physical character. 

The proposed mathematical apparatus is based on the effective (of polynomial complexity) classical algorithms, which give the dynamical picture of many particle quantum processes. Providing algorithms with the status of first principles makes possible to give exact formulations of the things which have no such formulations in standard quantum theory, for example, for the mechanism of concordance between unitary evolutions and measurements. It opens door for the regular treatment of quantum effects in the many body dynamics, which means the theoretical possibility of the complete acquisition of chemistry. The price of the proposed modification is recognizing a video film as a main form of the representation of results in theoretical physics, and the agreement to use analytically found values (for example, energies) only as the checkpoints for such a video film. This idea is not completely new, for example, many physicists actively use the packages of programs for numerical calculations and supplement the first principles of quantum theory by their own models of various processes. I represent the algorithmic view point in the radical form of the mathematical apparatus in order to show how wonderful consequences it gives to quantum mechanics and its applications to complex systems and how good it agrees with the spirit and traditions of physics. 

Quantum mechanics gives the most fundamental representation of the structure of matter; it forms the basic terms of our comprehension of the chemical and biological processes. Hence the question about its effective implementation to the many body systems (separate elementary particles, atoms, molecules) has the principle meaning. But just the structure of quantum theory contains elements which seriously complicate such implementation\footnote{Many body call it contra - intuitiveness of quantum physics.}. These elements are: “squeezing” of the wave packages, connected with the principle of uncertainty, interference of amplitudes, non locality, the existence of two principally different forms of evolution: unitary evolution and measurements. The discussion of these peculiarities has the long history, beginning from the famous discussion between Einstein and Bohr about the nature of quantum randomness. The contra intuitiveness of quantum description forced many body either to look for more “realistic” version of quantum theory or to lead these peculiarities to its limit(\cite{Bom}, \cite{Bel}, \cite{v.Neu}, \cite{Ev}, and also \cite{Me} and others.). In the past years with the appearance and active development of the idea of quantum computer (\cite{Fe}, and also \cite{Ben}, \cite{De}, etc.) we obtain the possibility of the different approach to this question that may be more fruitful. The point is that the peculiarities of quantum theory follow from its mathematical apparatus, which is based on mathematical analysis and Hilbert spaces. The tremendous superfluity of this mathematical apparatus remained unnoticed during all the time when quantum mechanics develops, because practically only analytical methods were applied (the solution of differential equations, matrices diagonalization) for one particle, whereas the many particle area of Hilbert formalism connected with the forming of new spaces by tensor product of existing remained without applications\footnote{The only exclusion are states which can be written as Schmidt expansion (see, for example \cite{VBB}), that is from the algebraic viewpoint the simple generalization of EPR pairs; we discuss them below.}. This superfluity has not interfered the development of quantum physics, because there was no experimental scheme verifying the adequateness of this many particle formalism. The situation radically changed with the appearance of quantum computer (QC) scheme. The quantum computer is the first principle prediction of quantum theory, e.g., it follows immediately from the formalism of Hilbert spaces. Consequently, if this formalism is adequate for many body quantum systems then QC must exist and will be built despite of serious difficulties lying on this way (see \cite{Va}). But it may happen that Hilbert formalism is not adequate in the area of many particles problems. The procedure of taking tensor product of one particle spaces of states may have the physical sense: it can reflect the real mechanism when two “samples” of these particles glue together and form the “sample” of new quantum particle. In this case the entanglement turns to be the physical resource and the realization of arbitrary complex states predicted by Hilbert formalism becomes impossible because it presumes the unacceptably huge (exponential) expenditure of this resource. This is what I call (physical) non adequateness of Hilbert many particle formalism\footnote{Sometimes it is expressed in more fuzzy terms: the productiveness of the device producing the more complicated entangled state must be multiplied to small factor (proportional to $e/c$) with every complication.}. This situation requires the elaboration of new mathematical apparatus for the description of quantum many particle phenomena. 

The nature of difficulties of QC technologies lies in the basis of quantum theory: in the existence of spontaneous decoherence leading to the reduction of quantum many particle states. The mechanism of such reduction lies beyond the standard quantum theory\footnote{There are many opinions about its possible construction: from the influence of gravitation (see \cite{Pe}), mental efforts of an observer (\cite{Me}, \cite{She} etc.), to the refusal of reduction at all (\cite{Ev}). We do not touch this theme here.}. It is important that the unitary evolutions (Shroedinger equation) without reduction procedure cannot give the complete picture of quantum dynamics\footnote{In contrast with classical physics where such a picture is possible.}. Hilbert formalism for many particles (even with QC) does not then leave us the hope to obtain once the description of Nature possessing the predicting power. By the predicting power we, of course, mean not a fate of single photon determining by quantum randomness, but the quantum mechanical probability predictions expended to the level of big ensembles of atoms and molecules (for example, to $10^{16}$ atoms that is sufficient for the representation of simplest bacteria). If we hope to obtain once such a description\footnote{Those who do not hope should probably change the science to the other area of activity.}, we should agree that something is not in the order with the Hilbert formalism itself, e.g., that the difficulties nest in the mathematical apparatus of quantum theory. This paper is devoted to the discussion what to do with this. 

\section{Why there is the problem with the mathematical apparatus of quantum theory and what can be done}

The systematic character of difficulties met by quantum theory witnesses that for the further development its mathematical apparatus must be changed\footnote{Somebody can feel horror from this idea, but this is only emotional reactions. The educational system and traditions force theorists to rely to the robustness of the conventional analytical and algebraic technique without doubts, but one must understand that the mathematics does not remain unchanged. For example, in nineteenth not many physicists could suppose what role the theory of continuous groups representations will play in physics.}. The limits of application of mathematical analysis to the processes connected with super large ($10^{20}$ and more) ensembles of small particles are well known. The value of differentials must be bigger than the size of elementary objects which compose the ensembles, because otherwise the analytical method will give the systematic mistake.  For example, the exact solution of the heat transfer equation gives the instantaneous speed of heat transfer that is impossible in practice. The analogous is true for Shroedinger equation which then cannot be transformed to the relativistic. The divergence from the exact solutions can serve as the indicator of grain of the considered system, that is applicable to the quantum mechanical amplitudes as well (the sequence from the supposition of grained amplitude is considered below). 

The other example of this kind is represented by quantum electrodynamics (QED). We consider the interaction between one electron and two photons: 1 and 2 (see \cite{Fe2}). Let at first the electron with impulse $p_1$ emit the photon 1 with impulse $k$, and then this electron emit the photon 2 with impulse $q$, and at last the electron absorb the photon 1 obtaining impulse $p_2$. Accordingly to the rules of QED the amplitude of such process is proportional to the integral 
\begin{equation}
\int\frac{1}{(p_2-k-m)(p_1-k-m)q^2k^2}d^4k,
\end{equation}
which diverges logarithmically. This difficulty is called ultraviolet divergence, and it cannot be eliminated by means of analytical methods\footnote{In physics it is resolved by the trick called renormalization (see \cite{BS}). Such tricks “correct” the undesirable presentations of analytical formalism features, but by no means can eliminate the fundamental defect of it.}. 

At last, the principle division of quantum evolutions to the unitary operators and measurements (reductions of the wave package) makes impossible to create the dynamical picture of quantum processes, because this description is made dependent from the free will of those who are authorized to observe.\footnote{The practical question: beginning from what quantity of particles the system must be considered as classical – has no answer in quantum theory. But this question is important for computer programs simulating chemistry because the picture of reactions depends on the exact answer to it. The standard criterion consisting of comparison of the action with Planck constant does not give the solution; because it merely transforms the problem to the question of the choice of the value of elementary time segment.}.

The possible solution of these problems is the modification of the mathematical apparatus of quantum theory. This modification can be based on the notion of algorithm. An algorithm is an instruction for the operation on a finite object expressed in the exact terms. The step-to step fulfillment of this instruction is called the computation. Formulas which are applied to the finite data types are examples of algorithms. But algorithms are much flexible than formulas, because the instructions can have not algebraic character. It is important what type of physical processes is required for the realization of the instructions. If there are classical processes, the algorithms are classical, if it needs the quantum processes, then we have deal with QC. Since in sense of computability QC is equivalent to classical computers, for the algorithmic approach considered in general form there is no difference to what extend QC can be made scalable. But if we want to use algorithms as the basis of mathematical apparatus, the way of physical realization of the algorithms is important. The main aim what QC has been proposed for is the simulation of many particle quantum physics. This is why the question about physical realization of algorithms is not the technical but the principal. This question can be reformulated as follows: is it possible to build the dynamical model of real quantum processes be means of classical algorithms, or the scalable QC is necessary for it ?\footnote{Of course, we do not intend to build classical algorithms operating with vectors in the space of exponential dimensionality. The point is that this exponential can be redundant for the description of Nature.} Factually, the only what is required from such QC is to realize quantum Fourier transform for any dimension of any particle from arbitrary big ensemble in quantum state (\cite{Za} and \cite{Wi}). This question is equivalent to the question: does a scalable QC exist at all. The existence of a scalable QC means the reality of such local physical processes which cannot be simulated by means of classical algorithms in the real time mode. There are the fast quantum algorithms (the most known of them – factoring of integers \cite{Sh}, the algorithm of fast quantum search \cite{Gr}; the more special algorithm can be found in \cite{Am}). These algorithms represent the test for checking for a given device: is it a QC or not.\footnote{This is the single universal test on QC and no other can serve as a complete criterion of it. The physical picture of the world thus depends on the complexity of algorithm !}. The existence of a scalable QC is then equivalent to the adequacy of many-particle Hilbert formalism. 

We note that all corollaries from quantum theory which are verified in experiments up to now can be derived by effective (requiring polynomial memory of the length of input) classical algorithms. At the same time Hilbert formalism asserts that for a system with $n$ particles all states from the space ${\cal H} = \bigotimes\limits_{j=1}^n{\cal H}_j$ are physically realizable. This space is the tensor product of one particle spaces and its dimensionality $dim({\cal H})=dim({\cal H}_1)dim({\cal H}_2)\ldots dim({\cal H}_n)$ grows exponentially with the growth of $n$. Hence, if Hilbert formalism is adequate then the classical computers are not appropriate for the simulation of physics. 

There is no yet clear experimental evidence that the scalable QC exists, despite of that the work goes in the different technological directions\footnote{Solid state quantum dots, Josephson junctions, ion traps; the last approach now becomes more promising (see \cite{SB}). The obtaining of entangled states with more than 10 particles is not the big problem now as well as the creation of 2 qubit gates. The problem consists in the checking of the real parameters of such devices that is reduced to the realization of fast quantum algorithms, or, as an intermediate solution to the checking of the probability distributions of qubit’s states on their corresponding to the quantum distributions.}. We must then account the possibility that the scalable QC will not be built, which means that we should create the program solutions for the simulation problems based on classical algorithms\footnote{And the simulators of limited QC as well. Classical simulators of quantum algorithms must be included to them as the autonomous element in the system of distributed computations and be realized on supercomputers. This scheme of simulation can serve as an indicator of success in the creation of (limited) QC.}. 

The following difficulty in the traditional mathematical apparatus is the impossibility to create the dynamical models of quantum systems. The dynamics of quantum system even of one particle substantially depends on the reduction of wave function (see \cite{Pe}) and quantum theory contains neither conditions of its realization, nor criterion how to consider a given system – as classical or as quantum. It results in that the dynamical quantum models are limited by the problems of scattering\footnote{The main application of them are the processes in colliders.} that are solved by $S$ matrix; it is absolutely not sufficient for the building of complex dynamical pictures, like the chemical reactions. Here just algorithms could serve as the main mathematical tool. 

The dynamical picture of evolution is the main aim of the proposed modification of the mathematical apparatus for quantum theory. Quantum physics in its traditional form in its nature cannot give the dynamical picture; it is designed for the solution of stationary problems only. At the same time, in my opinion, the future of physics is connected with the dynamical pictures. The example of biology says that there must be the real mechanisms resulting in the visible dynamical picture. I do not doubt that the mechanism of such a kind (but, of course, of absolutely other nature) exists on the physical level of description of the matter\footnote{Without such mechanisms the realization of great plans of the simulation of living cells is impossible. The simulation of molecular movements in a cell based on classical “balls and springs“ can give movies which can be fine for ordering of experimental facts only but have no prediction power, as fundamental theories, in particular, biological.}. The discovery of such mechanisms in physics seems to be the principal problem, and it cannot be solved (and even correctly formulated) by means of conventional Hilbert formalism for many quantum particles. This is why I think that the building of new mathematical apparatus for quantum theory based on effective classical algorithms is very important\footnote{Ideas of this kind were put forward by many theorists, for example, \cite{Ak}, \cite{Wo}, \cite{Oz}.}, whereas conventional algebraic and analytical technique will acquire the status of excellent heuristics despite of very precise corresponding with experimental data in the area of one particle dynamics.  

\section{What does it mean that algorithms acquire the status of mathematical apparatus}

Mathematical formalism plays the special role. The physical intuition is impossible without it\footnote{To be more precise, the intuition and mathematical apparatus are two sides of one thing.}. In particular it means that the property of physical theories which we call contra intuitiveness is factually the symptom of deeply lying defect of mathematical apparatus. The modification of mathematical apparatus is absolutely non trivial procedure which cannot be compared for example, with renormalization. The passage to the algorithmic formalism in quantum theory leads some principal consequences which we consider now. 

These corollaries are connected with the gradual computerization of physics and we then will use the computational terminology.\footnote{This question can cause perplexity among the part of specialists which could take  it as an encroachment on their individuality. Factually this touches the human individuality in the same degree as the mathematical apparatus of physics.}.

The main consequence: the form of representation of theoretical results and its interrelations with experiment will change. The main form of algorithmic description of the quantum evolutions is the “film” which pictures are prepared by the simulating algorithm. This form of representation follows from the nature of algorithms. If we are given an algorithm $A$ and some initial state of the object $O$ on which this algorithm must be fulfilled (input data) then in the general case there is no way to obtain the far result of work of $A$ on this state but sequential applications of the instruction corresponding to $A$\footnote{This fact remains true even if we have QC (see \cite{Oz2}).}. The single universal form in which the results of algorithms can be represented is its protocols, e.g., the sequential results of applications of the corresponding instructions. If we (roughly) associate the state of object $O$ with the states of simulated system, we conclude that the single universal form of the representation of the result of simulation is the corresponding “video film” reflected the dynamics of our system in time. 

This viewpoint is the roughening. The state of object $O$ on which our algorithm works can (and must) contain, beyond the main part, some additional elements that are called ancillary elements (ancilla). Physically, ancilla is the administrative part of the model which is invisible for a user who is the looker of the video film. Administrative part of the model bears the technical role but its existence is necessary for the preparation of the film. There is the physical interpretation of this property of the simulating algorithms. It is connected with quantum non locality and relativism \footnote{The bad agreement between the spirit of relativism and quantum non locality can follow from that the mechanisms of the both phenomena belongs to ancilla.}. Non locality of EPR photon states means that the video film cannot be prepared in the real time mode. For the simulation (on classical computer) of the results of measurements of two parts of the system in state $|00\rangle +|11\rangle$ we need the transfer of some information between the first and the second qubits and the time of this transfer must not be real, physical time\footnote{One could say that from the user’s viewpoint this information is transferred instantaneously. But this information belongs to ancilla, and user Alisa cannot then use this channel to transfer to the other user – Bob information created by Alisa. Relativism is not then violated.}. 

The visualization of quantum dynamics\footnote{Now it is considered mainly as the question of teaching physics (see, for example, \cite{Th}).}, then becomes the necessary part of theory and the main tool for the checking of hypothesis. In some sense the visualization acquires the higher level than the immediate application of algebraic or analytic methods for finding probability distributions in the elementary steps of evolution (for scattering problems) because the last methods are then considered as the necessary checkpoints for the building of right visual picture, not as the independent tasks. 

It is important that the simulating algorithm must be based on the limited and stated beforehand set of first principles, and must not contain any artifacts, which could fit the algorithm to the conditions of particular problem. This formulation depends on what we consider as the first principles for algorithms. We cannot simply take the basic principles of QED and claim them first principles because at first, their reduction to algorithms represent the substantial task, and at second, if we turn them to the instructions of algorithm, the formal implementation of such instructions\footnote{Mathematical apparatus presume only formal implementation.} gives infinite procedure because it meets the divergences. Factually, the reduction of QED to algorithms is fairly real task. But it is possible only if we introduce some computational tricks to the list of first principles\footnote{It can touch also the procedure of renormalization.}. These tricks represent the cut off Hilbert formalism in the sense that its computer realization will certainly give us some approximation to the real picture, whereas standard QED cannot give such approximation in principle. We consider some of such tricks below and discuss what tricks must be added for the creation of more full picture. 

We have to sacrifice some advantages of Hilbert formalism, in particular, the easy passages to other basis in quantum space of states. For example, the grain (quantum) of amplitude will depend on what basis (coordinate or impulse) we consider our system in. The coordinate basis will be separated for us. But we will show below how the consideration of state in impulse basis can be reduced to the coordinate basis (for photons). Some inequality of basis for representation of quantum systems is unavoidable in the reduction of standard formalism to algorithms. For example, in QED it is convenient to represent the states of charged particles in coordinate basis and the states of photons – in impulse basis, and it is reflected in the algorithmic reduction. 

The accepting of algorithms as the new mathematical apparatus means the agreement that the pure computational limitations acquire the status of fundamental physical laws. For example, the impossibility to reserve too large memory for the storage of complex states of many body system means that there are no such states at all\footnote{This gives the main method of approximate computations: to cut off the complexity of quantum states.}. For example, the decoherence will arise not as the result of environmental influence\footnote{Influence of environment is the conventional source of decoherence.}, but as the result of limitation on computer memory which can be reserved for the storage of complex quantum states of many particles. 

We can make this statement more precise, if we require that any real states of quantum $n$ particle system can be stored in the memory of the size $M(n)$, which depends on the number $n$ but not on the complexity of states. This requirement is connected with the question: what objects we consider as particles, and it is treated below. 

The main and universal form of description will be the representation of objects by visual images, not by formulas. This viewpoint is not conventional. We show that the standard quantum mechanics with the entanglement of Schmidt type completely agrees with this representation. The problem arises with QC. If we orient to the visualization, it means that we can use only effective classical algorithms, e.g., algorithms which time of the work is limited from above by some polynomial of the input word\footnote{Better if there are algorithms with linear complexity. Just these algorithms are considered in the work below.}. This required the radical reduction of Hilbert formalism. The visualization of QC in its scalable form can be impossible just because of existence of the fast quantum algorithms\footnote{One would separate the “visual” part of the model where only Schmidt type of entanglement can be, and “non visible” part, where a scalable QC can be (see below). This is the good solution for the indication of success in the going experiments on QC. But this solution is not appropriate for the role of mathematical apparatus of quantum theory because of violation of integrity.}, which work cannot be visualized because the essential intermediate steps of it cannot be adequately represented in the visual terms. In any case, the final arbiter here is only one: the practical building of scalable QC; this is the only way to reject the algorithmic approach. 

\section{Amplitude quantum and Born rule}

Many of equations in mathematical physics (heat transfer, diffusion, oscillations) describing the classical systems dynamics result from the passage to limit in the dynamical problems of huge quantities of small bodies (quanta of the matter). Correspondingly, the area of application of such equations is limited by the finite sizes of these bodies. These equations resulted from the more fundamental laws or mechanisms of interaction (for example, the equation of oscillations follows from Hooke law). In QED such mechanism is described by Feynman diagrams for fundamental processes (\cite{Fe2}). The fundamental processes diagrams facilitate the algorithmic reduction of QED but yet are not the final result of this reduction because they allow the operations with infinitesimals. Algorithms require the complete transfer to the operations with the finite objects\footnote{The wide spread mistake that this would limit the possibilities of the theory has the same origin that the fear of algorithms. Nobody can operate with the infinite objects as well as with non algorithmic procedures. The question is only in that some elements of algorithms can be inaccessible for us (as the administrative part in quantum model). As for the internal beauty, it is not less in the world of algorithms than in its part: in the world of formulas. But algorithms have one principle advantage: the possibility of visualization. The mathematics which does not permit the visualization can rely on the deduction only, but it is the precarious basis, in view of Goedel theorem about incompleteness. The reckless usage of such mathematics is similar to walking on a thin ice.}.
The method of collective behavior described below gives the possible form of the algorithmization \footnote{See also \cite{Oz}, \cite{SO})}. We now consider one aspect of this form of algorithmization – the quantization of amplitudes. 

We give the explanation of Born rule based on the conception of amplitude quantum\footnote{The similar reasoning contains in \cite{Zu}, but here Born rule follows from the general algorithmic concept without additional suppositions.}.

The consideration of quantum evolutions from the viewpoint of Hilbert formalism of many particles gives the states of the form

\begin{equation}
|\Psi\rangle=\sum\limits_j\la_j|e_j\rangle,
\end{equation}
where the summing extends to the unlimited set of basic states of the system $|e_j\rangle$. Algorithmic approach requires to cut off this row to the finite sum by elimination of all summands with coefficients $\la_j$, which modules are less than some constant threshold $\e$. We call this procedure the reduction, and agree to fulfill the reduction on any state which we meet in evolution. Such reduced sum contains no more than $1/\e^2$ summands. Let $N$ be the number of basic states for one particle. We then can agree that $\e=\frac{1}{\sqrt{N}}$. The state thus will have the form

\begin{equation}
|\Psi\rangle=\sum\limits_{j=1}^N\la_j|e_j\rangle,
\label{limstate}
\end{equation}
where some summands can be zeroes.  

We call the constant $\e$ amplitude quantum. 

We now show how the reduction leads to the Born rule for the finding of quantum probabilities. 
For this we reduce the finding of probability to obtain some certain basic state $A$ in the measurement of state $\Psi$ to the application of classical rule 
$$
p(A)=\frac{N_{suc}}{N_{tot}}
$$
where $N_{suc}$ is the total number of successful events (e.g., such elementary events for which the event $A$ is realized), $N_{tot}$ the total number of elementary events. We define the set of elementary events and establish the correspondence between elementary events and the basic states of the measured system. We call an elementary event such a state of extended system (measured system  + measuring apparatus) which module of amplitude in the given quantum state equals $\e$. The set of elementary events will then depend on a quantum state of extended system. 

Let $|\Psi_j\rangle$ denote basic states of measured system, and $|\Phi_j\rangle$ denote basic states of measuring device (which can be, for example, the eye of observer), we obtain in the instant of contact between these two subsystems the state of the form
\begin{equation}
\sum\limits_j\la_j|\Psi_j\rangle\bigotimes |\Phi_j\rangle
\label{meas}
\end{equation}

Now, in view that the measuring device is very massive in comparison with the measured object, when trying to describe its states we must divide states from (\ref{meas}) to the sum of $l_j$ basic states (we must account states of all nuclei and electrons containing in the measuring device, all photons emitted and absorbed by it, etc.). If even in the instant of contact we have the state $|\Phi_j\rangle$, the evolution transforms it very fast to the state $|\Phi'_j\rangle =\sum\limits_{k=1}^{l_j}\mu_{j,k}|\phi_{j,k}\rangle$, where the numbers $l_j$ increase rapidly in time until the amplitudes modules achieve $\e$ - when they will become nulls. Consequently, all modules of amplitudes $\mu_{j,k}$ must be considered as approximately equal. If we substitute the expression for $|\Phi'_j\rangle$ instead of $|\Phi_j\rangle$ in (\ref{meas}), the amplitudes of states $\phi_{j,k}$ become equal to $\frac{\la_j}{\sqrt{l_j}}$ because the quantum evolution is unitary.  

We agreed to fulfill the reduction, that is the elimination of summands $\phi_{j,k}$ which module of amplitude is too small. Since the time frame when the division to the huge number of summands is very short, in the computations it means that we divide each summand in (\ref{meas}) to $l_j$ new summands so that all newly arisen amplitudes have modules close to amplitude quantum and approximately equal, that makes all the states equivalent before the reduction and it makes possible to apply the classical urn scheme. The quantity $l_j$ of summands with the first multiplier  $|\Psi_j\rangle$, that is the total number of successful elementary events, is proportional to $|\la_j|^2$, and if only one summands survive in the reduction, we obtain Born rule for quantum probability.  

The probability space then depends on the choice of wave function $|\Psi\rangle$. We consider, factually, the conditional probabilities to obtain the results of measurements provided the system is in state $|\Psi\rangle$. 

The explanation of Born rule we give is based only on our definition of the wave function reduction as the cancellation of small amplitudes. This reduction is fulfilled at each step of the unitary evolution simulation, because otherwise the simulation would be impossible at all. Here the specific of measurement comparatively to the unitary evolution is only quantitative: the measurement is the moment when our system falls in contact with the massive object which can be called “environment” that causes the division of summands in (\ref{meas}) to the large number of new summands. Beyond this natural supposition we used only the stability of the wave function norm which follows from Shroedinger equation. Hence, in our explanation of Born rule we used nothing from outside of standard quantum mechanics but the reduction of wave function concluding in the cancellation of small amplitudes. Just this reduction procedure transforms the set of Feynman paths into the classical trajectory for an object with the big action (see \cite{FH}). 
Decoherence is treated as the forming of entangled states of the form (\ref{meas}) with the environment, which means that we do not separate it from the especial measuring of the considered system. Born rule and the irreversible corruption of states called decoherence is thus the consequences of the existence of amplitude quantum. Factually, decoherence and measurements follow from the severe limitation of classical memory of the simulating computer. The algorithmic first principles thus acquire purely physical form\footnote{The quantitative estimation of the speed of decoherence can be obtained from the simulation with sequential reductions. The only difficulty is how to account all states of our system, depending of all the degrees of freedom, which we often do not know exactly. Here the hierarchical description of complex system (see below) can help.}.

We note that in the derivation of Born rule we used only the entanglement of (generalized) Schmidt type of the form (\ref{meas}), which generalizes EPR type of entanglement. 

The simple treatment of Born rule in terms of amplitude quantum is not arbitrary. The concretization of amplitude quantum – the method of collective behavior, introduced below, makes possible to build the dynamics of many particle evolution practically. 

\section{Method of collective behavior}

The method of collective behavior (swarm method) is the simplest and most evident way for the algorithmization of quantum theory. The passage from the single particle to the swarm of its samples seems to be the easiest way to overcome contra intuitiveness featured to quantum theory. With some additional suppositions swarm method gives the algorithm of simulation of the dynamics with linear complexity of the number of particles. These additional suppositions lie in the framework of the basic idea of algorithmic approach – the limitation of the memory and time for the simulation\footnote{The question is only how severe these limitations are. Here we discuss the case of the most severe limitations – the linear growth of computational resources. The constant is so that in practical simulation we can operate with all entangled states of no more than 50 qubits.}. The supposition is that we should treat as realizable only states of the system $S=S_1\cap S_2,\ S_1\sup S_2=\emptyset$ of the form $|\Psi_S\rangle=\sum\limits_j\la_j|\Psi^j_{S_1}\rangle\bigotimes|\Psi^j_{S_2}\rangle$, where $|\Psi^1_{S_1}, |\Psi^2_{S_1},\ldots$ and $|\Psi^1_{S_2}, |\Psi^2_{S_2}$ are orthonormal basis in space of states of subsystems $S_1$ and $S_2$ correspondingly, and each of these basic states has the same form, with some depth of nesting. Factually, this is one particle states, where the entanglement is distributed among the particles sequentially nested in each other (see below). Moreover, we can make our approach completely scalable only if the sets $\{|\Psi^j_{S_1}\}_j$ and $\{|\Psi^j_{S_2}\}_j$ belong to coordinate (for charged particles) or impulse (in case of photons) basis of the space of states. 

\subsection{Swarm representation of particles and fields}

We represent any quantum particle in the form of swarm (set) of classical particles 
\begin{equation}
\label{swarm}
s=\{ s_1,s_2,\ldots,s_k\}
\end{equation}
where each element $s_j$ is called a sample of the considered quantum particle, and has the definite spatial-time coordinates $t(s_j),\bar x(s_j)$ in the configuration space-time corresponding to this particle, and some auxiliary parameters $\a (s_j),\b (s_j),\g (s_j),\ldots$.

A swarm can be stored in the form of a field. We divide the configuration space to cells $C_i$, and in each cell we find the total number of samples $s_j$, for which $\a (s_j)=\a_0$. It gives the natural number expressing the intensity of scalar field $F_{\a=\a_0}$, corresponding to the chosen value $\a_0$ of the parameter $\a$. If $\a_1,\a_2,\ldots,\a_l$ are some values of $\a$, the set of corresponding fields $F_{\a=\a_i},\ i=1,2,\ldots,l$ is called the vector field and its intensity in each point is the vector from the corresponding natural numbers\footnote{The appropriate normalization gives the vector from the real numbers.}.
The storage of intensity gives the exponential economy comparatively with the storage of massive of samples. But for the representation of dynamics we need just the notion of samples. If the parameters of quantum particle (or some set of particles) are not so interest for us than the intensity of it, we call the field the representation of this particle (particles) in the form of massive of its intensity values in the whole configuration space\footnote{E.g., the difference between fields and particles is only methodical. For example, scalar photons is more convenient to represent as a unique field, but the vector potential – in the form of separate photons, because the polarization of scalar photons is parallel to their impulses and thus it bears no additional information over their total number, whereas the vector photon polarization is orthogonal to the impulse vector, and it bears the information determining electrical and magnetic field induced by these photons.}. 
In the case when some quantum particle is concentrated in one cell of our division of configuration space we call it the classical particle. The classicality of particles is thus determined by the division of space-time. 

Swarm representation of a system of $n$ particles and $l$ fields is the set of vectors of the form
\begin{equation}
\label{gen_swarm}
\bar s_{par} =\{ s^1,s^2,\ldots,s^n\},\ \ \bar \g=\{\g_1,\g_2,\ldots,\g_l\} , 
\end{equation}
where $\bar s_{par}$ are the swarms of samples of particles, $\bar \g$ are the swarms of samples of fields. Evolution of the representation (\ref{gen_swarm}) in time is determined by the set ${\cal T}$ of rules for particle transformations, each of which has the form
\begin{equation}
 v_{j_1},v_{j_2},\ldots,v_{j_p }\ar v_{k_1},v_{k_2},\ldots,v_{k_q},
\end{equation}
where $ v_{j_1},v_{j_2},\ldots,v_{j_p }$ и $ v_{k_1},v_{k_2},\ldots,v_{k_q}$ are the sets of initial and resulted samples of particles and fields which can belong to the different swarms but satisfy the following requirement of locality. The coordinates of samples from the both groups can be obtained one from the other by the simple rule which we call the rule of local correspondence. In the simplest case this rule consists in that the coordinates of them must coincide, e.g., the interaction must be point wise (the action of a scalar potential). The more complex case is that these coordinates can differ only on some small value $\e>0$ (free flight of particles). The more complex form of local correspondence rule is for the passage from the coordinate representation to impulse. Here the rule says that the samples from one group must lie approximately on the equal distances from one another with interval $d$, and are close to some fixed line, where $d$ and the position of the line is determined by the coordinates and parameters of initial samples (energy conservation in the emission and absorption of photon by the charged particle). 

Here are the main forms of the rules for particle transformations.
\begin{equation}
\begin{array}{lll}
1).\ &v_p,v’_p&\longleftarrow\ar ,\ \ v,v’\in\{ s,g\},\\
2).\ &g_p&\ar g_q,\ \\
3).\ &s_p&\ar s_p,g_q,\ \\
4).\ &g_p&\ar (g_p,) s_q,\ \\
5).\ &s_{j_1},s_{j_2},\ldots,s_{j_p }&\ar g_{k_1},g_{k_2},\ldots,g_{k_p },\ \\
6).\ &g_{j_1},g_{j_2},\ldots,s_{j_p }&\ar s_{k_1},s_{k_2},\ldots,s_{k_p },\\
7).\ &s_p,s_{p’}&\longleftarrow\ar s_q.
\end{array}
\end{equation}
These transformations determine correspondingly: the normalizing of particles and fields, diffusion of fields, emission of virtual photon samples by the samples of charged particles, emission of charged particle samples by the samples of virtual scalar photons, emission of vector photon samples by the samples of charged particles, emission of charged particle samples by the samples of virtual vector photons (photon absorption), the forming and the decay of entangled states of Schmidt type. The free flight of charged quantum particles as well as the action of Coulomb potential is determined by the transformations of types 3) and 4), the spreading of Coulomb field and the field of vector photons – by 2), interaction between charged particles and vector photons - by 5) and 6). The normalizing 1) is auxiliary and is applied in every instant. 

Every rule of transformation of particles is fulfilled with the corresponding intensity (the probability to apply the rule to the appropriate set of samples). The intensities of the application of rules are constants taken from the experiments\footnote{One could try to derive them, but here the more complex things must be accounted, as the relativity delays.}.
 
\subsection{Swarm representation of Shroedinger equation}

Shroedinger equation for the wave function $\Psi(x,t)$ has the form

\begin{equation}
i\dot\Psi=-\D\Psi+V\Psi .
\label{shr}
\end{equation}
We represent the wave function in the form 
$$
\Psi(x,t)=\Psi^r(x,t)+i\Psi^i(x,t),
$$
where $\Psi^r(x,t),\ \Psi^i(x,t)$ are its real and imaginary parts. The equation (\ref{shr}) can be then written in the form of system of two equations
\begin{equation}
\begin{array}{lll}
&\dot\Psi^r&=-\D\Psi^i+V\Psi^i,\\
&\dot\Psi^i&=\D\Psi^r-V\Psi^r.
\end{array}
\label{}
\end{equation}

We establish the necessary connection between Shroedinger equation and the equation of diffusion. The equation of diffusion has the form

\begin{equation}
\rho_1\dot u=div(p\ grad\ u)-qu+F,
\label{diff}
\end{equation}
where $u(x,t)$ is the concentration (density) of particles, $\rho_1,p,q,F$ parameters depending on $x,t$, that determines the density of environment, diffusion rate, absorption rate and the source of particles intensity correspondingly. The positive absorption rate means that all the particles are eliminated in this point with the intensity $q$, and negative means that they are created in this point with intensity $|q|$. We agree that $F=0,$ а $\rho_1, p$ have the unit values and thus only the absorption rate $q$ has the substantial sense; it is proportional to the potential energy on Shroedinger equation. The equation (\ref{diff}) follows immediately from Nernst law (see (\cite{Vl}) 
for finding of the stream of particles through the element of surface $dS$:
\begin{equation}
dQ=-p\frac{\partial u}{\partial\bar n}dS.
\label{ner}
\end{equation}

To reduce the equation (\ref{shr}) to some version of the diffusion equation we consider the swarm of samples of our particle. We divide these samples to two types: real ($r$) and imaginary ($i$), and each of these types divide to two subtypes: positive (+) and negative (-). We then obtain the subdivision of all samples of the same particle to four types which members we denote as: $ \a^{+,r}_j,\ \a^{+,i}_j, \ \a^{-,r}_j,\ \a^{-,i}_j,$ where $j$ denotes the number of the sample. For the description of stationary states only one type of samples would suffice, because stationary states are determined by the density of wave function. For the description of dynamics we could manage with only 2 types of samples: real and imaginary, but it is convenient to have four types. We divide the configuration space on cubes so that $D(x,t)$ denotes the cube containing the point $(x,t)$. 
The total number of samples of the swarm $\bar s$ of the same type occurred in some cube $D(x,t)$, is denoted by $s^{\sigma ,\eta}(x,t)$, where $\sigma\in\{ +,-\},\ \eta\in\{ r,i\}$. We agree that the speeds of all samples are distributed uniformly and independently of the type. Since we are going to represent the evolution of wave function as the chain of sequential diffusions, we have to get rid of signs in equations and for this the introduced types of samples must be used, where the swarm approximation of wave function is always found by the formula 
\begin{equation}
\Psi(x,t)_{s}=s^{+,r}-s^{-,r}+i(s^{+,i}-s^{-,i}).
\label{sign}
\end{equation}
This equation does not determine the division to the positive and negative parts uniquelly, but only within the addition of the same constant to the both parts. 

We then has within the normalizing the approximate equalities
\begin{equation}
\begin{array}{lll}
&\Psi^r(x,t)&\approx s^{+,r}(x,t)-s^{-,r}(x,t),\\
&\Psi^i(x,t)&\approx s^{+,i}(x,t)-s^{-,i}(x,t),
\end{array}
\end{equation}

This gives the system, equivalent for Shroedinger equation:
\begin{equation}
\begin{array}{lll}
&\dot s^{+,r}(x,t)&=\D s^{-,i}(x,t)+V(x,t)s^{+,i}(x,t),\\
&\dot s^{-,r}(x,t)&=\D s^{+,i}(x,t)+V(x,t) s^{-,i}(x,t),\\
&\dot s^{+,i}(x,t)&=\D s^{+,r}(x,t)+V(x,t) s^{-,r}(x,t),\\
&\dot s^{-,i}(x,t)&=\D s^{-,r}(x,t)+V(x,t) s^{+,r}(x,t).
\end{array}
\label{swa}
\end{equation}

We enumerate the types $(+,r),(+,i),(-,r),(-,i)$ by the natural numbers $1,2,3,4$ correspondingly everywhere including indices, and apply to them arithmetic operations in $Z/4Z$. We already have no one density $u$, but the vector- column $\bar u$, which components $u_j\ j=1,2,3,4$ are the densities of the samples of types $j$. For our aim the rule of type transformation must correspond to the going around the beginning of coordinates in some direction, e.g., the rule must be cyclic: $1\ar 2\ar 3\ar 4\ar 1$. 
The system (\ref{swa}) is thus equivalent to the following equation 
\begin{equation}
\dot\bar u=\G (\D u-q\bar u),
\label{diff_trans}
\end{equation}
 where the matrix $\G$, expressing the law of type transformation and the matrix $g$ inverting the sign of type have the form
$$
\G=
\left(
\begin{array}{lllll}
&0&0&0&1\\
&1&0&0&0\\
&0&1&0&0\\
&0&0&1&0
\end{array}
\right) ,\ \ \ 
g=
\left(
\begin{array}{lllll}
&0&0&1&0\\
&0&0&0&1\\
&1&0&0&0\\
&0&1&0&0
\end{array}
\right) .
$$

We now have to determine the swarm behavior giving the solution of equation (\ref{diff_trans}). Here it is impossible to manage with only reactions of type transformation accordingly to the cyclic rule, because in this case the matrix $\G$ would contain the equal and nonzero diagonal elements which all are zeroes in $\G$. This is why we need to introduce the samples of connected photons to the swarm behavior. 
 
Let we are given 4 types of the samples of particle: 1,2,3,4, possessing the same dynamical properties, and also the same types of the samples of connected photons. We denote the samples of particle and connected photons by $\a^j$ and $\g^j$ correspondingly. Let every sample of the charged particle of the type $j=1,2,3,4$ emits with the steady rate the samples of connected photons of the same type. These samples of connected photons move by the diffusion and in time frame $\d t$ they convert to the samples of the same particle but of the type $j+1$. 
The type transformation thus goes not directly but through the samples of connected photons. The main which we need from the samples of connected photons is that their diffusion rate must be much larger than of the samples of particle: $p_{phot}\gg p_{part}$. We could agree that the sample of particles do not move themselves at all. Just this requirement allow to suppress the diagonal elements in the matrix of generalized diffusion and thus to obtain the needed matrix $\G$.

We then consider two areas $D_1$ and $D_2$ with the common border through which the diffusion goes. Here due our agreement the samples of photons diffuse much faster than the samples of particle.  
In the change of the density of each type in the small area in the time frame $\d t$ two kind of samples of this type make their deposits: a) newly formed samples resulted from the photon samples emitted in this area or penetrated from the neighboring areas, and b) penetrated by the immediate diffusion. Due to our agreement about the values of diffusion rate the deposit b) will be much less than a), and we will neglect the immediate diffusion of samples of particle. Then applying to our situation the formula (\ref{ner}), we obtain that the flow of samples of the type $j$ in the unit time frame through the border is 
\begin{equation}
dQ=-p_{phot}\frac{\partial u^{j-1}}{\partial\bar n}dS,
\label{ner1}
\end{equation}
where $u^{j-1}$ is the density of samples of particle belonging to the parent type. The potential $V$ is treated as the rate of creation or annihilation of samples of any type. Applying the reasoning from the derivation of diffusion equation we come to the equation (\ref{diff_trans}). 

The introduced method of description of quantum evolution includes classical dynamics of samples of particles. It is sufficient to have the quantity of the connected photon samples $A(|s^{r,+}-s^{r,-}|^2+|s^{i,+}-s^{i,-}|^2)^{1/2}$, where $A$ is some constant independent of the point of the configuration space that prevent the useless storage in the mutually canceling parts of wave function in the memory. Let the carrier of wave function be concentrated in the area $D_0$. The phase distribution of the form $\phi(x)=С \bar x\bar p$ then gives the same effect that the movement of the swarm with the speed proportional to $\bar p$ along the vector $\bar p$. Correspondingly, we can change the rule for the transformation of the connected photon samples to the samples of particles, using the mean speed $v_{av}(x,t)$ of the samples of particles emitting the connected photon samples in the point $x,t$. The rule defined below says that the connected photon samples of type $j$ transform to the samples of particles of type $j+1$ in the instant $t+\d t$, here $\d t$ - constant. 

The new rule will be: the connected photon samples transform to the samples of particle of the type corresponding to the amplitude distribution $exp(\d tip^2/m)$, where $\d t$ is the arbitrary positive number in the segment $(0,\d t_0)$, where the impulse $p=v_{av}m$ is found from the mean speed of the particle samples which emit the connected photon samples. It means that for the fixed $\d t$ in this segment the fraction of the type $j+k, \ k=0,1,2,3$ equals to the corresponding amplitude in the exponential $exp(\d tip^2/m)$. This new rule can be realized by the transformation of type 5), and the reverse operation – by 6). 

The new rule does not effect to the description of swarm dynamics if the initial amplitude distribution has the form $exp(ipx)$ and $V=0$. Here the movement will be uniform and straight with the impulse $p$. In the other cases the influence of this rule is determined by the value of $\d t_0$: the less it is the less the influence will be. If we fix $\d t_0$, we can find the influence of the potential $V$. Let $\D \Psi (\bar\D x)$ be the divergence between the value of wave function found by the old and new rules for the transformation of photon samples to the samples of particle in the point $x+\bar\D x$. The value $\D \Psi (\bar\D x)$ is then proportional to $\sin (\bar{grad}\ V,\bar\D x)$. The divergence will be then maximal if the direction of the photon samples moving is orthogonal to $\bar{grad}\ V$.
It substantiates the following agreement. We assume that a sample of connected photon transforms to the free photon sample if it spreads to the distance more than some limit $\D x_0$ from the point of emission along $\bar\D x$, so that $|\sin (\bar{grad}V,\D x)|>\e_0$, e.g., along the direction close to the normal for the gradient of the potential in which the considered particle moves. This agreement is not completely precise because we have not determined the constants $\e_0,\ \D x_0$, but it can be reformulated in terms of QED if we introduce the vector of polarization of photon which is always orthogonal to its impulse. Our agreement will then express in terms of swarm the rule for finding of the amplitude of the fundamental process: emission (absorption amplitude is complex conjugate) of the photon by the charged particle (see (\cite{Fe2})). This amplitude is imaginary and it is proportional to $i(p_1+p_2)(\bar \varepsilon , p_1+p_2)$, where $ p_1, p_2$ - are impulses of the particle before and after the emission (absorption), $p_1=p_2+q$, where $q$ - is the photon impulse, $\bar \varepsilon$ - its polarization. This rule is equivalent to Dirac relativistic equation for spineless particle.  We can realize it by our rules of the type transformation of the types 5) and 6). If we then introduce the photon spirality, which determines its magnetic field, we can show that Maxwell equations can be derived from it\footnote{We consider non relativistic theory but in the swarm form also the relativistic equations can be represented regularly, for example, Dirac equation for free electron. For this we must not ignore the own diffusion of the particle samples as we did it above. Moreover, we could introduce the spins of particles as the limits of momentums of impulses of the corresponding field if the size of spatial cell reaches the minimal size. We then can represent the influence of the spin of particle in the given point on the spirality of emitted (absorbed) photon sample and obtain, for example, spin-spin interaction of electrons and nuclei, and spin-orbit interaction.}. 

\subsection{Representation of Coulomb field in the method of collective behavior}

The spreading of scalar field in the method of collective behavior goes on diffusion mechanism, e.g., as follows. Each sample of the field $g$ with the probability $p$ remains where it was, and with the probability $\frac{1}{6}(1-p)$ shifts to each neighboring cell. It is realized by the rule 2) for the transformation of particles. This mechanism is analogous to Coulomb flow of wave package of the free particle. We suppose that the samples of Coulomb field of the charged particle are emitted by all samples of this particle with the constant intensity proportional to its charge $e$. Let the coefficient of diffusion for the field $p$ is much larger than this coefficient for the virtual photon samples which determine the spreading of the wave package of charged particle (the field, induced by the particle must exist in the areas where there is no samples of this particle). Then in the short time frame when the density of the samples of particle has not changed substantially, the field reaches its stationary state. The intensity of stationary state of the field $\phi$ created in the point $x_0$ is the solution of the equation $\Delta\phi (x)=\delta(x_0)\phi (x)$, e.g., it is Green function  of Laplace operator $\Delta$, determining the diffusion. This function in the space $R^3$ is proportional to $\frac{e}{r}$, 
$r=|x-x_0|$. Coulomb potential can be thus obtained by the diffusion method. Analogously we can build in the swarm representation any other potential determined by a rule of the form of cell automata, e.g., by a rule of local type. We can also use the non local rules for the field construction in the case when there is the simple algorithm for the selection of the spatial cells participating in this rule. For example, we can obtain the local interaction in the impulse space by the selection of fixed quantity of points located in the same straight line with the equal intervals. Here the total number of the required steps will be always proportional to the total numbers of elements in the division of configuration space. 

The advantages of swarm representation of Coulomb field are the following. To find the state of quantum system with $n$ particles on one step of evolution it requires $O(nN)$ operations, where $N$ is the number of elements of the division of configuration space. It follows immediately from the locality of all considered interactions. If we suppose the instantaneous spreading of Coulomb potential we would have to consider the interactions for all the pairs of particles ($O(n^2)$), and for each such pair – the interaction of all pairs of their samples ($O(N^2)$), that gives the total number of operations $O((nN)^2)$. The finite speed of spreading of the filed determining interaction thus gives the lowest possible complexity of the simulation needed for the scalability. 

We note that the estimation of the complexity from above is valid for the case when we do not use the transformation of the particles (rule 7)) and all the particles remain unchanged. In this case our method of simulation can represent only non entangled states. The possibility to represent entangled states is discussed below. 

\section{Scalability of the collective behavior method}

From the formal viewpoint the main drawback of Hilbert formalism is the lack of scalability. We treat the scalability as the possibility to add new particles to the considered system which does not cause the revision of all the computational work which has been already fulfilled for previous system. The best conditions of scalability is the situation when the complexity of simulation of one step of evolution of the system with $n$ particles is $O(n)$\footnote{This, in particular, means that the precision of the one particle description must not depend on $n$, e.g., the variety of one particle states must not suffer from the existence of entanglement. Just this is the most difficult point in the creation of a quantum computer independently of its technology.}.   
Since the main expenses of the time and memory in the standard Hilbert formalism owe to the work with the entangled states, the advantage of the collective behavior method in sense of scalability is connected mainly with the representation of entangled states. We formulate the conditions under which the collective behavior method would be completely scalable and show how they can be checked in experiments. These conditions can also serve as the indicators of intermediate success in the building of quantum processors with big total number of qubits because its experimental checking is much easier than the realization of the fast quantum algorithms\footnote{The absolute indicator of success in the building of QC is the reliably working QC with 60 logical qubits (and the same quantity of ancilla) on which one could solve by Grover method the search problem inaccessible for the classical computers. Taking into account quantum error correction, the number of logical qubits must be multiplied to $10^5$. Our criterion is applicable yet to 3 qubits.}.

\subsection{Entangled states as the form of particles}

The single mean for the description of entanglement in the collective behavior method is the forming of new particles from the existing ones by the operation 7). For the preserving of advantages of swarm method it is undesirable to extend its signature, hence in the description of entangled states we will use only the operation 7). 

The quantum system dynamics can lead to the situation when two particles glue and begin to behave as the single particle. The example is the joining of one electron and one proton to the molecule of hydrogen (with the emission of photon), or (with the addition of neutrino) to neutron. If we ignore the internal excited levels of such composed system and account its components states by the rules applicable to two independent particles we can roughly represent the states of such composed particle as $|00\rangle+|11\rangle$, where the first qubit denotes the spatial position of the electron, the second – of the proton. This state can be easily distinguished from the mixture $|00\rangle$ and $|11\rangle$ with equal probabilities, for example, measuring impulses of the both parts of the composed particle. We thus have the real entangled state of two quantum particles. This state can be represented in terms of collective behavior, using the rule 7) of the particle transformations: $e+p\ar H$. To ensure our requirement that the total numbers of values of parameters of any particle must be much less than the total number of elements of the configuration space division, we must consider the electron in this system in the stationary state (for example, $1s$) as occupying only one position in the configuration space of the joint particle. Only then we could consider the hydrogen atom as one quantum particle which (samples) can interfere. In the opposite case (for example, if dependently of the position of atom the electron occupies the excited levels) no interference will be, e.g., the particle ($H$) will not be one, but many ($H_{1s}, H_{2s}, H_{2p}$, etc.). Analogously can be described the complex molecules, cooper pairs of electrons in superconductors, pairs of entangled photons or pairs “charged particle+ photon”, entangled ions in the trap. Here in the case of photons the configuration state for them is always impulse. The internal states of the composed parts can be quantum stationary (or non stationary) states or even classical states; but it is important that these states do not depend on the position of joint particle in its configuration space (e.g., the joint particle must interfere with itself). We consider, for example, by Born- Oppengeimer scheme, the rolling molecule of hydrogen $H_2$. If it rolls independently of its center of mass position, we can apply the rule 7) for it: $H+H\ar H_2$ and consider it as the joint particle in the framework of the collective behavior method. 

We consider Hilbert representation of the system of particles which in turn consist of more small particles, etc. We agree that every particle of the level $k-1$ is located in the center of mass of forming it particles of the level $k$, and in the group of these particles the determining of coordinates of all but one determines the coordinates of the last one (in the system of center of mass). Let $r_1,r_2,\ldots,r_n$ be coordinates of all particles partially ordered by nesting so that the particles of the more depth level of nesting has the much number, and the particles of the same level of nesting follow in the random order. 

We apply the qubit representation of the wave functions $|\Psi(\bar r)\rangle$ in the form 
\begin{equation}
\sum\limits_{\bar r}\la_{\bar r}|\bar r\rangle
\label{wave_function}
\end{equation}
where $\bar r$ is the binary arithmetic notation of the values of coordinates of all particles containing in the considered system. Let $n$ be the length of this list. Here the value of wave function $\Psi(\bar r)$ is proportional to $\la_{\bar r}$. We consider the natural lexicographic order on the list $\bar r$, corresponding to the one particle case, but our consideration will be the general.\footnote{Representation of wave functions in the form (\ref{wave_function}) is more convenient than the traditional form $\Psi(\bar r)$, because the last form can be ambiguous the two different objects can be denoted by it: the wave function and its value in the concrete point $\bar r$ (so to distinguish these two senses physicists often write integrals with delta functions).} We denote by $\bar r_k$ the initial segment of the sequence $\bar r$ of the length $k$, where $r_k$ is the $k$-th element of this sequence. If we do not use the upper indices, we can agree that the $r_k$ represent only one qubit – this makes our notations easier. Any wave function of the form (\ref{wave_function}) can be represented as
\begin{equation}
|\Psi\rangle=\sum\limits_{r_1}\left(\la_{\bar r_1}|r_1\rangle\bigotimes\sum\limits_{r_2}\left(\la_{\bar r_2}|r_2\rangle\bigotimes\ldots\bigotimes\sum\limits_{r_n}\la_{\bar r_n}|\bar r_n\rangle\right)\ldots\right)
\label{hie}
\end{equation}
For this we can simply take all $\la_{\bar r_j}$ equal 1 for $j=1,2,\ldots,n-1$, and for $j=n$ take them equal $\la_{\bar r}$ from the formula (\ref{wave_function}). 

If we consider the fixed value of $j$, the amplitude distribution $\la_{\bar r_j}$ can be treated as some wave function; we suppose that it is normalized. Let all the functions $\la_{\bar r_j}$ really depend not on all the list $\bar r_{j}$, but on the coordinates $r_{j-p},r_{j-p+1},\ldots,r_{j}$ only. Then we call the set of such states the class of the depth $p$. The class ${\cal P}_0$ of the states of zero depth in the simple generalization of Schmidt states to the case of hierarchical system of nested particles. It is formed by the states for which the distributions of the samples oa any particles in the coordinate system of the embracing sample $s$ of the particle of the lesser level of nesting do not depend on the coordinates of $s$. The class of states of the depth $p$ is denoted by ${\cal P}_p$.  

The class of states ${\cal P}_0$ admits the natural swarm representation. We include each amplitude distribution $\la_{\bar r_j}$ to the parameters of the corresponding particle of the level $j-1$. For example, the wave function of the system “electron+proton” with the motionless center of mass is the joint parameter of the particle “hydrogen atom”. The state (\ref{hie}) can be then obtained by the sequential operations of the type 7) of the form 
\begin{equation}
s_j,s_k\ar s_q,
\label{reaction}
\end{equation}
 where $s_j,\ s_k$ are the particles containing to the particle $s_q$. 

The states of the class ${\cal P}_k$ can be reduced to the states of the class ${\cal P}_{k-1}$, by the introduction of new particles which differ from the existing in their amplitude distribution in the coordinate system connected with their center of mass. Every state can be then reduced to ${\cal P}_0$ \footnote{It can be not optimal because the quantity of the types of particles can grow too fast. In any case it is necessary to preserve the linear growth of complexity, whereas the criteria on which we distinguish the different particles from the different states of the same particle can depend on the method of visualization.}. 

For example, we consider the atom of lithium Li, in its ground state. It consists of: nucleus, one electron with spin up, one electron with spin down in state $1s$ and one electron with spin up in state $2s$ (we ignore the deformation of states resulted from the Coulomb interaction between the electrons). This atom is the joint particle which samples can interfere with each other. The full state of such a particle $Li_0$ belongs to the class ${\cal P}$, and is an example of entangled state. We suppose that this atom interacts with the vector photon and can pass to the excited state $Li_1$. Such an excited atom will be considered as the different particle, with the different constituents: for example the third electron will be not the particle $2s$, but the particle $3p$ (the other electrons also will be the particles of the other types, but we neglect it). The long and simultaneous existence of the samples of the two different types of particles: $Li_0$ and $Li_1$ is impossible because it results in the super expenditure of the memory comparatively with the initial situation when $Li_0$ and the photon were independent\footnote{These different samples of  $Li_0$ and $Li_1$ cannot interfere. For example, if the initial $Li_0$ interferes on two slits and the photon is oriented to one of these slits, then after the absorption the interference picture disappears.}. 
The fast choice between two alternatives: $Li_0$ + photon flied away, or $Li_1$ without photons is unavoidable for the method of collective behavior. The reason is that otherwise we would not have enough memory for the storing of the more and more complex states connected with the further behavior of the virtual photon and virtual atom which will be in the intermediate states. This choice is fulfilled by the probability mechanism as it was shown in the section about Born rule. 

In terms of the transformations of particles of the form (\ref{reaction}) it means that this reaction either goes for all the pairs of the form $ s_j,s_k$ (and the types of samples $s_j,s_k$ are contained as the parameters in the type of particle $s_q$), or does not go to anything. Hence we accept the following principle of stability of swarm: the samples of the swarm corresponding to the same particle behave in the transformations of the type 7) equally. It is impossible to distinguish the samples of one swarm to two different swarms. The two swarms can result from one swarm can happen only if every sample divides to two different samples of the different particles in the transformation of the type 7) reverse to (\ref{reaction}). Inside of the same swarm all the samples must be treated as identical because they must interfere with each other. Hence, if we detect the divergence from the identity leading to the impossibility of interference (as in the example with $Li_0$ and $Li_1$), we must introduce the new particle\footnote{If we allow the possibility to divide the swarm to two and forming the two real particle from one, then with the unlimited size of memory we could write any state of the form (\ref{hie}) through the transformation of particles  in the form of such states of the zero depth of influence. I claim that this form of notation of the entangled states is optimal on the expenditure of classical memory. In the other words, the most valuable part of quantum interference lies inside of one swarm. If the Hilbert form of representation of the dynamics leads to the increasing of the depth of influence, it always leads to the forming of new physical particles (e.g., ensembles which much effective can be treated as whole particles). For example, when the electron flies at proton the spreading of the electron wave function $\Psi_e$ means physically the transformation of the far located branches of $\Psi_e$ to photons. The certain physical sense expressed by the particle transformation stays behind every “over expenditure of the classical memory”.}.   

The description of entangled states in the method of collective behavior is thus the principle constriction of the Hilbert formalism for many body quantum systems. The checking of the fact of belonging of the given many body state to the class ${\cal P}_k$ is reduced to the comparison of the statistics for this class with the experimental statistics of such states which can be regularly obtained in big quantities and with the high accuracy. It gives, in principle, the indicator of the success in the experiments on the quantum processors, as well as the checking of our basic supposition about the validity of collective behavior method. The rougher indicator checks only entanglement of states and its degree\footnote{In principle it is possible the brief existence of the intermediate states, for example, of the depth 1,2, or more. Such states do not belong to the class ${\cal P}_0$, but can be represented in the swarm form if we increase the quantity of the types of particles and over expend memory in the process of the reaction. But if we suppose that the swarm approach is scalable, then this over expenditure means simple the redistribution of the computational resource and the rougher description of some other process. In any cases the life time of these more complex states must be much lesser than of the states from ${\cal P}_0$.}. 

The entangled states of the type ${\cal P}_0$ arise in the case if the movements of the pair (or bigger number) of particles are somehow synchronized. It can take place if there is some attracting potential between the particles which is induced by themselves (as for atoms in vacuum) or their interaction with the other particles (ions in traps, cooper pairs). The outside influence can be essential, for example, in the phase transitions where the new particles arise from the molecular clusters, which interference properties can influence to the picture of the dynamics. The existence of entangled states is connected not only with the electrodynamics interactions. The transformation of the particles in the nuclear interactions and the effects resulted from the entanglement of the form ${\cal P}_k$ between nucleons and atomic structures can play the important role in the dynamics observed in experiments. The role of such effects cannot be determined using the standard quantum formalism for many particles, but we hope to go farer by the method of collective behavior.\footnote{One more sign of the perspective ness of the collective behavior method is the accuracy of the computation of stationary states of electron systems in atoms and molecules by the diffusion Monte Carlo method which in the stationary analog of the collective behavior method. Such computations give the most exact agreement with the experimental data.}.

\subsection{Interpretation of the identity of particles in the collective behavior method}

The spaces of occupational numbers corresponding to boson or fermion symmetry of the ensemble of identicl particles represent the important part of standard Hilbert formalism. It means that in the algebraic operations with the vector of state $|\Psi\rangle$ of the system of $n$ identical particles which separate states have the form $|\psi_j\rangle=\sum_k\la_j^{k_j}|k_j\rangle ,\ j=1,2,\ldots,n$ in case of independent states, inspead of $|\Psi\rangle=\bigotimes_j |\psi_j\rangle $ we should always write  
\begin{equation}
|\Psi\rangle=A\sum\limits_{\pi_1,\pi_2,\ldots,\pi_N}D_{j,r}(\la_j^{k_{r}})|\pi_1,\pi_2,\ldots,\pi_N \rangle ,
\label{fock}
\end{equation}
where $|\pi_1,\pi_2,\ldots,\pi_N \rangle$ are the states of system which means that in any $l$-th element of the division of the configuration state for one particle $l=1,2,\ldots,N$ there are $\pi_l$ particles, where the sum spreads to such summands for which for all $s=1,2,\ldots,N$ and $k$ $\sum_{r:\ k_{r}=s}=\pi_s$. Here for a fixed $k$ for fermions $D_{j,r}$ is the determinant, and for bosons - permanent of the matrix $(\la_j^{k_{r}})_{j,r=1,2\ldots,n}$. Such states are treated as non entangled in the Fock space of the occupational numbers ${\cal F}_n$ for $n$ particles, and by the linear combinations with them we can obtain arbitrary states in this space. The states $|\pi_1,\pi_2,\ldots,\pi_N \rangle$ for an orthonormal basis of ${\cal F}_n$. The coefficient $A$ is determined from the normalizing; in the case of  orthonormal $|\psi_j\rangle$ it equals $\frac{1}{\sqrt{n!}}$. We consider the case of identical fermions. 

In the collective behavior method any state from Fock space ${\cal F}_n$ is represented as the set of $n$ swarms $S_1,S_2,\ldots,S_n$, which fill the mutually nonoverlapping areas of the (one particle configuration) space $D_1,D_2,\ldots, D_n$. The wave function (\ref{fock}) will be then the sum of functions which differ only by the renaming of their argumants and signs, and each of them has the swarm representation. This agreement does not touch the computation of the measured magnitudes; for example, the computation of the energies of stationary states by such functions must give the same result as in the standard Fock space, etc. 

\section{Spacio-temporal aspect of the collective behavior method}

We have yet noted the possibility to represent the relativistic wave equation (of the second order on the time) which we can try to obtain in the swarm approach if do not ignore the diffusion of the samples of charged particle. Here we discuss the other aspect of the collective behavior method touching the relativism. This aspect is connected with the known feature of the relativistic physics – the impossibility of the objective (independent of the observer) separation the past from the future. The collective behavior method has this feature as well. The dynamical picture of evolution sometimes cannot be build by the unilateral algorithm: from the past to the future. For the building of the next picture of the video film the specification of some details of the previous pictures can be necessary.  This necessity does not lead to the logical paradoxes by the same reason as the analogous feature of the relativistic physics: it does not violate the cause-and-effect chain of events in the space-time but only makes indispensable (in contrast to the non relativistic theory) the common consideration of the space and time. 

We consider a wave function in the relativistic (station-temporal) qubit representation 
\begin{equation}
\Psi=\sum\limits_{\bar r,\ t}\la_{\bar r,\ t}|\bar r,\ t\rangle
\end{equation}
where $\bar r$ and $t$ takes values from all points of the space and time correspondingly. The normalization of wave function is the dynamical process (see above) and hence it is normalized only for the values of $t$ filling some small but nonzero segment $t,t+\Delta t$ for each value of the time $t$\footnote{This is the reason of the difficulty of the normalization of wave functions for relativistic objects, as photons. For this the value $\Delta t$ must be fairly large for that the photon energy can be determined sufficiently exactly: the uncertainty relation gives $|\Delta t\ \Delta E|=h$.}. The norm of wave function cannot be thus exactly preserved for the very small time frames $\Delta t$. Just for such time frames the relativistic effects play the role that is connected with the relativity of the order of the events in the time. These effects consists in the reactions of the type: $photon\ \ar \ photon \ +\ particle\ + \ antiparticle$, in which anti particles participate. Anti particles are the particles which move in the reverse direction in the time. In QED antiparticles arise in the attempt to reduce QED to the fundamental processes (see 
\cite{Fe}) due to the symmetry of Dirac equation. We show how anti particles arise in the collective behavior method. 

Non relativistic description of evolution which we dealt with above is based on that the cause-and-effect connection of events and the physical time were treated as the same. In particular, we ignored the self diffusion of the charged particles samples and regarded only their transformations through the samples of connected photons, which gave Shroedinger equation. If the speed of samples of particles will be of the same order as the photon samples, then this supposition loses its force. We consider the process when the electron $e_0$ flies into the hollow $H$ through the slit $E_0$, and can leave this hollow through one of the exit slits: $E_1,E_2,\ldots, E_k$. Let the speed of electron be such that in the time of one fundamental process (emission of photon or its absorption) the electron can overpass the whole hollow and leave it through one of exit slits. If we consider this process as the sequential scatterings following in the order of physical time, we deal with the reactions of transformations of electrons and photons which fill some tree with the initial vertex $E_0$. Our aim will be the determining of the amplitudes of coming out of the electron through the slits $E_1,E_2,\ldots ,E_k$. We will fulfill computations in this tree until its upper branches reach all exit vertexes, corresponding to $E_1,E_2,\ldots ,E_k$. This requires the time of the order exponential of the (conditional) high of the tree $exp(h)$, where $h$ is the conditional distance from the beginning of the tree to the most far exit vertex.

We consider the other method of computations, where we suppose that the electron $e_0$ flies in the hollow through the slit $E_0$, and simultaneously from the hollow through $E_1$ the other electron $e_1$ comes out, which was born with the positron $e^1$ from the photon $p_1$. This positron then passes through the chain of sequential transformations and annihilates with the initial electron   $e_0$, transforming to the virtual photon $p_1$. Here we also must suppose that the photon can move back in the time, e.g., it is its own anti particle. In this method the cause-and-effect connection does not correspond to the physical time. But here we have two trees growing towards each other and the computation ends when its upper branches meet\footnote{We suppose that the vertexes of trees are disposed in the points of the division points in the configuration space corresponding to the hollow.}. The complexity of computation will have the order $exp(h/2)$, because the high of trees in their finite position will be twice lower than in the first case. E.g., the complexity of the second method of computations is of the order of square root of the complexity of first method. This is the heuristic argument for the introduction of anti particles in the swarm method. 

The relativistic effects have then the natural representation in terms of the collective behavior. It points to the possibility of the building of the algorithmic quantum formalism on the base of the collective behavior. 

\section{Conclusion}

We have given some arguments for the necessity of the modification of the mathematical apparatus of quantum mechanics of many body systems. The main argument is the principal impossibility to build the dynamical model for many particles in the framework of standard Hilbert formalism. This depreciates in the practical sense the great advantage of quantum theory consisting in the stability of quantum trajectories in the unitary evolutions. The replacement of the quantum Hilbert many-particle formalism by the more convenient formal tool for the representation of the many body dynamics is desirable. The effective classical algorithms are proposed to be such formalism. The unavoidable consequences of it: the fundamental character of the visual representation of quantum dynamics, the treatment of the limitations to the time and memory as the physical laws, including of decoherence to the general description of the dynamics, removing the observer from the quantum theory. Born rule follows from this representation of quantum physics. The price of modernization is the refusal from the idea of a scalable quantum computer. 

We also propose the concrete form of the algorithmization for QED: the collective behavior method. It is described in details on the example of standard Shroedinger equation, and the idea of the generalization to the interaction with photons is given. We also propose the simple description of entangled states based on the joining of several particles to one new particle. It was shown how to indicate the success in the experiments with quantum process using this representation of entangled states. The collective behavior method is completely scalable – it requires the time depending linearly from the quantity of particles in the considered system. Its regular application makes possible to build the good approximation of quantum dynamics for systems of many particles (thousands) that is impossible in the framework of standard formalism.

\end{document}